\documentclass[pra,aps,twocolumn,superscriptaddress,showpacs]{revtex4}
\usepackage{epsfig,amsmath}

\begin{document}

\title{A scheme for unconventional geometric quantum computation in cavity
QED}
\author{Xun-Li Feng}
\affiliation{Department of Physics, National University of Singapore, 2 Science Drive 3,
Singapore 117542}
\affiliation{State Key Laboratory of High Field Laser Physics, Shanghai Institute of
Optics and Fine Mechanics, Chinese Academy of Sciences, Shanghai 201800, P.
R. China}
\author{Zisheng Wang}
\affiliation{Department of Physics, National University of Singapore, 2 Science Drive 3,
Singapore 117542}
\author{Chunfeng Wu}
\affiliation{Department of Physics, National University of Singapore, 2 Science Drive 3,
Singapore 117542}
\affiliation{Institute of Theoretical Physics, Northeast Normal University, Changchun
130024, P. R. China}
\author{L. C. Kwek}
\affiliation{Department of Physics, National University of Singapore, 2 Science Drive 3,
Singapore 117542}
\affiliation{Nanyang Technological University, National Institute of Education, 1 Nanyang
Walk, Singapore 637616}
\author{C. H. Oh}
\affiliation{Department of Physics, National University of Singapore, 2 Science Drive 3,
Singapore 117542}

\begin{abstract}
We present a scheme for implementing the unconventional geometric two-qubit
phase gate with nonzero dynamical phase by using the two-channel Raman
interaction of two atoms in a cavity. We show that the dynamical phase
acquired in a cyclic evolution is proportional to the geometric phase
acquired in the same cyclic evolution, hence the the total phase possesses
the same geometric features as the geometric phase. In our scheme the atomic
excited state is adiabatically eliminated and the operation of the proposed
logic gate involves only in the metastable states of the atom and hence is
not affected by spontaneous emission.
\end{abstract}

\maketitle

\section{Introduction}

\label{1} Quantum computation employs the principle of coherent
superposition and quantum entanglement to solve certain problems, such as
factoring large integers and searching data in an array, much faster than a
classical computer \cite{Bennett}. The basic building blocks of a quantum
computer are quantum logic gates. It was shown that any quantum computation
can be reduced to a sequence of two classes of quantum gates, namely,
universal two-qubit logic gates and one-qubit local operations \cite%
{Universal}. The standard paradigm of quantum computation is the
dynamical one where the local interactions between the qubits are
controlled in such a way so that one can enact a sequence of quantum
gates. On the other hand, it has been recognized that the quantum
gate operations can also be implemented through the geometric
effects on the wave function of the systems, this is the so-called
geometric quantum computation \cite{Zanardi}. Compared with the
dynamical gates, the geometric quantum computation possesses
practical advantages. It is well known that geometric phases depend
only on some global geometric features, and do not depend on the
details of the path, the time spent, the driving Hamiltonian, and
the initial and final states of the evolution \cite{berryph}.
Therefore the geometric quantum computation is largely insensitive
to local inaccuracies and fluctuations, and thus provides us a
possible way to achieve fault-tolerant quantum gates.

In the implementation of geometric quantum computation, one
practical question we usually meet is how to remove or avoid the
dynamical phases since geometric phases are generally accompanied by
dynamical ones which are not robust against local inaccuracies and
fluctuations. To this end one simple method is to choose the dark
states as qubit space, thus the dynamical phase is always zero
\cite{duan01}. Another general method is to let the evolution be
dragged by the Hamiltonian along several special closed loops, then
the dynamical phases accumulated in different loops may be
canceled, with the geometric phases being added \cite{ekert00,falci00,WangXB}%
. This is the so-called multi-loop scheme.

The geometric quantum computation which is based on the cancelation of
dynamical phases is referred to as conventional geometric quantum
computation. Correspondingly several schemes have been presented recently to
realize the so-called unconventional geometric quantum computation \cite%
{leibfried03,unzhu,zheng,Feng}. The central idea of the unconventional
geometric quantum computation is that for certain quantum evolution of a
quantum system of interest one can implement fault-tolerant quantum
computation by using the total phase accumulated in the evolution if it
depends only on global geometric features of the evolution. In comparison
with conventional geometric gates, unconventional geometric gates do not
require additional operations to cancel the dynamical phases and thus
simplify the realization operations. Schemes for implementing the
unconventional geometric gate have been proposed in trapped ion systems \cite%
{leibfried03,unzhu} and in cavity QED systems\cite{zheng,Feng}. In the
schemes of cavity QED systems\cite{zheng,Feng}, the excited states are
utilized as the computational bases, thus the spontaneous emission cannot be
avoided in such schemes.

In this paper we make use of the two-channel Raman interaction in cavity QED
\cite{two-channel,walther} to realize the unconventional geometric gate. In
our scheme the atomic excited states are adiabatically eliminated and never
excited during the quantum gate construction, therefore atomic spontaneous
emission can be avoided in our scheme.

\section{Theoretical model of two-channel Raman coupling in a cavity\ }

\label{2}
\begin{figure}
\begin{center}
\epsfig{figure=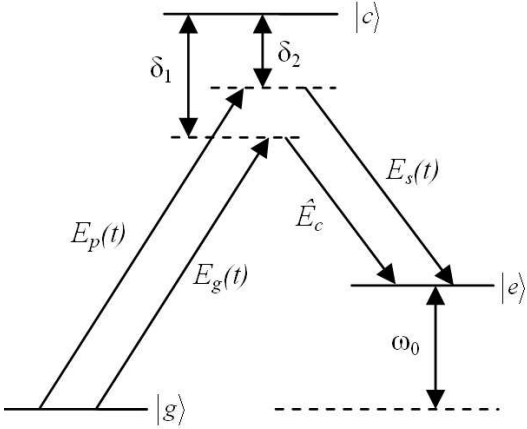,width=0.4\textwidth}
\end{center}
\caption{The two-channel Raman transition diagram. The detunings
$\delta_1$ and $\delta_2$ are assumed to be sufficiently large, so
that the excited state $|c\rangle$ can be eliminated.} \label{fig1}
\end{figure}

We consider two identical three-level atoms in $\Lambda $%
-configuration placed in a high-\textit{Q }cavity. The level
structure
of the atoms is shown \ in Fig. \ref{fig1}, where $|e_{i}\rangle $, $|g_{i}\rangle $ (%
$i=1,2$) are metastable states and $|c_{i}\rangle $ is an excited
state. The
transitions $|c_{i}\rangle \leftrightarrow |g_{i}\rangle $ and $%
|c_{i}\rangle \leftrightarrow |e_{i}\rangle $ are supposed to be
dipole-allowed. Each atom can be off-resonantly excited via two Raman
channels by laser fields and the cavity mode. One channel is excited by two
classical external fields $E_{p}(t)$ and $E_{s}(t)$\ with the frequencies $%
\omega _{p}$ and $\omega _{s},$\ respectively. The second channel contains a
classical external field $E_{g}(t)$ with a frequency $\omega _{g}$ and a
quantized cavity field of frequency $\omega _{c}$. Both channels are assumed
to satisfy the usual Raman resonance, that is, $\omega _{p}-$ $\omega
_{s}=\omega _{g}-$ $\omega _{c}=\omega _{0},$ where $\omega _{0}$ is the
energy difference between levels $|e\rangle $ and $|g\rangle $. The electric
field operator of the cavity mode can be expressed as $\widehat{E}%
_{c}=k(a+a^{\dagger })\overrightarrow{e}_{\lambda }$, where $a$ and $%
a^{\dagger }$ are respectively the annihilation and creation operators, $%
\overrightarrow{e}_{\lambda }$ is the polarization vector and $k$ is
a constant determined by the quantization volume. In the case that
the detunings $\delta _{1}$ and $\delta _{2}$ are sufficiently
large, the atomic excited state $|c\rangle $ can be adiabatically
eliminated and we can obtain an effective Hamiltonian \ref{fig1}.
The detunings $\protect\delta _{1}$ and $\protect\delta _{2}$\ are
assumed to be sufficiently large, so that the excited state
$|c\rangle $ can be eliminated.
\begin{equation}
H(t)=\sum_{j=1}^{2}\left[ r(t)\sigma _{j}^{+}+r^{\ast }(t)\sigma _{j}^{-}%
\right] +\sum_{j=1}^{2}\left[ g(t)a\sigma _{j}^{+}+g^{\ast
}(t)a^{\dagger }\sigma _{j}^{-}\right] , \label{h1}
\end{equation}%
where $\sigma ^{+}=|e\rangle \langle g|$ and $\sigma ^{-}=|g\rangle \langle
e|$ are atomic operators, $r(t)$ and $g(t)$ are respectively the effective
classical and quantum couplings and they take the following form%
\begin{eqnarray}
r(t) &=&-\frac{\left[ \langle c|\overrightarrow{d}|g\rangle \cdot
\overrightarrow{E}_{p}(t)\right] \left[ \langle e|\overrightarrow{d}%
|c\rangle \cdot \overrightarrow{E}_{s}^{\ast }(t)\right] }{\delta _{1}}, \\
g(t) &=&-\frac{\left[ \langle c|\overrightarrow{d}|g\rangle \cdot
\overrightarrow{E}_{g}(t)\right] \left[ \langle e|\overrightarrow{d}%
|c\rangle \cdot k\overrightarrow{e}_{\lambda }\right] }{\delta
_{2}}.
\end{eqnarray}%
Here $\langle i|\overrightarrow{d}|j\rangle $ ($i,j=g,e,c$) denote
the atomic dipole matrix elements. From above equations it is easy
to note that the effective coupling parameters $r(t)$ and $g(t)$ can
be controlled by adjusting the driving light fields. Based on such a
feature of the Hamiltonian (\ref{h1}) a scheme to generate arbitrary
quantum states of the cavity fields was proposed \cite{two-channel}.

Now we further make the following transformation on the Hamiltonian (\ref{h1})%
\begin{equation}
H_{I}=\exp \left( iH_{0}t\right) \sum_{j=1}^{2}\left[ g(t)a\sigma
_{j}^{+}+g^{\ast }(t)a^{\dagger }\sigma _{j}^{-}\right] \exp \left(
-iH_{0}t\right) ,
\end{equation}%
where $H_{0}=\sum_{j=1}^{2}\left[ r(t)\sigma _{j}^{+}+r^{\ast }(t)\sigma
_{j}^{-}\right] $. For simplicity, we assume $r(t)$ is real, then $%
H_{0}=r(t)\sum_{j=1}^{2}\left[ \sigma _{j}^{+}+\sigma _{j}^{-}\right] .$
After a simple calculation we obtain%
\begin{eqnarray}
H_{I} &=&\frac{1}{2}\sum_{j=1}^{2}\left[ |+\rangle _{jj}\langle +|-|-\rangle
_{jj}\langle -|+e^{i2r(t)t}|+\rangle _{jj}\langle -|\right.  \notag \\
&&\left. -e^{-i2r(t)t}|-\rangle _{jj}\langle +|\right]
g(t)a+H.c.,\label{e5}
\end{eqnarray}%
where $|\pm \rangle _{j}=(|g\rangle _{j}\pm |e\rangle
_{j})/\sqrt{2}$ are eigenstates of $\sigma _{j}^{x}=\sigma
_{j}^{+}+\sigma _{j}^{-}$ with eigenvalues $\pm 1$, respectively. In
the strong effective classical driving regime $r(t)\gg |g|,$ the
terms in Eq. (\ref{e5}) which oscillate with high frequencies can be
eliminated in the rotating-wave approximation, and Eq. (\ref{e5})
can thus be simplified as
\begin{eqnarray}
H_{\mathrm{eff}} &=&\frac{1}{2}\sum_{j=1}^{2}\left( |+\rangle _{jj}\langle
+|-|-\rangle _{jj}\langle -|\right) \left[ g(t)a+g^{\ast }(t)a^{\dagger }%
\right] ,  \nonumber \\
&=&\frac{1}{2}\left[ g(t)a+g^{\ast }(t)a^{\dagger }\right] \left(
\sigma _{1}^{x}+\sigma _{2}^{x}\right) .\label{e6}
\end{eqnarray}%
Similar Hamiltonians have been derived in the strongly driving
Jaynes-Cummings model \cite{StrongJCM} and the two-channel Raman interaction
in cavity QED \cite{walther}. In comparison with the trapped-ion model
proposed to realize the unconventional geometric quantum computation \cite%
{unzhu}, the Hamiltonian in our model is of a similar form to that of the
trapped-ion model. In the next section we will show how to construct an
unconventional geometric two-qubit phase gate based on the above Hamiltonian.

\section{Unconventional geometric two-qubit phase gate}

\label{3} We choose the eigenstates of $\sigma _{j}^{x}(j=1,2)$, that is, $%
|\pm \rangle _{j}=(|g\rangle _{j}\pm |e\rangle _{j})/\sqrt{2},$ as
the computational basis, so that the Hamiltonian (\ref{e6}) will not
give rise to any population changes in such a computational basis
when the system is governed by the Hamiltonian (\ref{e6}). In the
computational basis \{$|+\rangle _{1}|+\rangle _{2}$, $|+\rangle
_{1}|-\rangle _{2}$, $|-\rangle _{1}|+\rangle _{2}$, $|-\rangle
_{1}|-\rangle _{2}$\} the Hamiltonian (\ref{e6}) is diagonal and
takes the form
\begin{equation}
H_{\mathrm{eff}}=\frac{1}{2}\left[ g(t)a+g^{\ast }(t)a^{\dagger
}\right] \times \mathrm{diag}[\lambda _{++},\lambda _{+-},\lambda
_{-+},\lambda _{--}],\label{e7}
\end{equation}%
where $\lambda _{kl}$ $(k,l=+,-)$ are the eigenvales of $\left( \sigma
_{1}^{x}+\sigma _{2}^{x}\right) $ and $\lambda _{++}=-\lambda _{--}=2$, $%
\lambda _{+-}=\lambda _{-+}=0.$ The time evolution matrix $U(t)$ is thus
diagonal
\begin{equation}
U(t)=\mathrm{diag}\left[ U_{++}(t),1,1,U_{--}(t)\right] ,
\end{equation}%
where the diagonal matrix elements $U_{kl}(t)$ can be derived from
Eq. (\ref{e7}),
\begin{eqnarray}
U_{kl}(t) &=&\hat{T}e^{-i\frac{1}{2}\lambda _{kl}\int_{0}^{t}\left[
g(t)a+g^{\ast }(t)a^{\dagger }\right] dt},  \notag \\
&=&\lim_{N\rightarrow \infty }\prod_{n=1}^{N}e^{-i\frac{1}{2}\lambda _{kl}%
\left[ g(t_{n})a+g^{\ast }(t_{n})a^{\dagger }\right] \Delta t},  \notag \\
&=&\lim_{N\rightarrow \infty }\prod_{n=1}^{N}D\left[ \Delta \alpha
_{kl}(t_{n})\right] ,\label{e9}
\end{eqnarray}%
where $\hat{T}$ is the time ordering operator, $\Delta t=t/N$ is the time
interval, and $\Delta \alpha _{kl}(t_{n})=-i\frac{1}{2}\lambda _{kl}g^{\ast
}(t_{n})\Delta t,$ $D\left( \alpha \right) $ is the displacement operator
which takes the form $D(\alpha )=\mathrm{\exp }[\alpha a^{\dagger }-\alpha
^{\ast }a].$ The displacement operators satisfy the following relation
\begin{equation*}
D(\alpha )D(\beta )=e^{i\mathrm{Im}(\alpha \beta ^{\ast })}D(\alpha +\beta ).
\end{equation*}%
Based on the above formula, the Eq. (\ref{e9}) can be further
simplified as
\begin{equation}
U_{kl}(t)=e^{i\gamma _{kl}}D(\int_{c}d\alpha _{kl}),
\end{equation}%
with $\gamma _{kl}={\rm Im}\left( \int_{c}\alpha _{kl}^{\ast
}d\alpha _{kl}\right) $ and
\begin{equation}
d\alpha _{kl}=-i\frac{1}{2}\lambda _{kl}g^{\ast }(t)dt.\label{e11}
\end{equation}%
For a closed path $c$, $U_{kl}(t)=e^{i\gamma _{kl}}D(0)=e^{i\gamma
_{kl}}.$ Here $\gamma _{kl}$ is the total phase acquired by the
state $|k\rangle
_{1}|l\rangle _{2}$ $(k,l=+,-)$  in the cyclic evolution from $t=0$ to $t=T$%
. The total phase $\gamma _{kl}$ consists of two parts, one part is
geometric phase $\gamma _{kl}^{g}$, and the other part is the
dynamical phase $\gamma _{kl}^{d}$ \cite{berryph}. According to the
coherent-state
path integral methods \cite{Kuratsuji,Hillery,unzhu}, the geometric phase $%
\gamma _{kl}^{g}$ and the dynamical phase $\gamma _{kl}^{d}$ can be
calculated in the following way
\begin{eqnarray}
\gamma _{kl}^{g} &=&\frac{i}{2}\int_{0}^{T}(\alpha _{kl}^{\ast }\dot{\alpha}%
_{kl}-\dot{\alpha}_{kl}^{\ast }\alpha _{kl})dt,\label{e12}   \\
\gamma _{kl}^{d} &=&-\int_{0}^{T}H_{kl}(\alpha _{kl}^{\ast },\alpha
_{kl};t)dt,\label{e13}
\end{eqnarray}%
with
\begin{equation}
H_{kl}(\alpha _{kl}^{\ast },\alpha _{kl};t)=\langle \alpha
_{kl}(t)|H_{kl}(t)|\alpha _{kl}(t)\rangle .\label{e14}
\end{equation}%
From Eq. (\ref{e11}) we obtain
\begin{equation}
\alpha _{kl}(t)=-\frac{i}{2}\lambda _{kl}\int_{0}^{t}g^{\ast }(\tau
)d\tau .\label{e15}
\end{equation}%
Substituting Eq. (\ref{e7}) and Eq. (\ref{e15}) into Eq.
(\ref{e14}), we get
\begin{eqnarray}
&&H_{kl}(\alpha _{kl}^{\ast },\alpha _{kl};t)  \notag \\
&=&-\frac{i}{4}\lambda _{kl}^{2}\left[ g(t)\int_{0}^{t}g^{\ast }(\tau )d\tau
-g^{\ast }(t)\int_{0}^{t}g(\tau )d\tau \right] ,  \notag \\
&=&-\frac{i}{4}\lambda _{kl}^{2}G(t),\label{e16}
\end{eqnarray}%
here for the sake of simplicity we have set
$G(t)=g(t)\int_{0}^{t}g^{\ast }(\tau )d\tau -g^{\ast
}(t)\int_{0}^{t}g(\tau )d\tau $. With Eq. (\ref{e15}) and
Eq. (\ref{e16}), the geometric phase $\gamma _{kl}^{g}$ and the dynamical phase $%
\gamma _{kl}^{d}$ can be calculated according to the formulas (\ref{e12}) and (\ref{e13}),%
\begin{eqnarray}
\gamma _{kl}^{g} &=&-\frac{i}{8}\lambda _{kl}^{2}\int_{0}^{T}G(t)dt, \\
\gamma _{kl}^{d} &=&\frac{i}{4}\lambda _{kl}^{2}\int_{0}^{T}G(t)dt,
\end{eqnarray}%
and the total phase is given by
\begin{equation}
\gamma _{kl}=\gamma _{kl}^{g}+\gamma _{kl}^{d}=\frac{i}{8}\lambda
_{kl}^{2}\int_{0}^{T}G(t)dt.
\end{equation}%
Comparing the above three equations, we have,%
\begin{equation}
\gamma _{kl}=\frac{1}{2}\gamma _{kl}^{d}=-\gamma _{kl}^{g}.
\end{equation}%
The relations between the total phase $\gamma _{kl}$, the dynamical phase $%
\gamma _{kl}^{d}$ and the geometric phase $\gamma _{kl}^{g}$
indicate that in the system examined here the total phase $\gamma
_{kl}$ and the dynamical phase $\gamma _{kl}^{d}$ possess the global
geometric features as the the geometric phase $\gamma _{kl}^{g}$
does. Therefore the cyclic evolution
\begin{equation}
U(T)=\mathrm{diag}\left[ e^{i\gamma },1,1,e^{i\gamma }\right]
\end{equation}%
with $\gamma =\frac{i}{2}\int_{0}^{T}G(t)dt$ is a two-qubit phase gate
operation which is robust against some lacal inaccuracies and fluctuations,
this gate is nontrivial if $\gamma \neq 2n\pi .$ As described in the
preceding section, the effective coupling parameter $g(t)$ can be controlled
by adjusting the driving light field, so that the cyclic evolution and
certain total phase $\gamma =\frac{i}{2}\int_{0}^{T}G(t)dt$ can be achieved.

It is worth noting that in our scheme the atomic excited state is
adiabatically eliminated and never populated. The quantum phase gate
operation only involves atomic metastable states , therefore the
effect of the spontaneous emission can be ignored.

\section{Conclusions}

\label{4} In this paper, we present a scheme for implementing the
unconventional geometric two-qubit phase gate with nonzero dynamical
phase by using the two-channel Raman interaction of two atoms in a
cavity. We show that the dynamical phase acquired in a cyclic
evolution is proportional to the geometric phase acquired in the
same cyclic evolution, hence the the total phase possesses the same
geometric features as the geometric phase does. In our scheme the
atomic excited state is adiabatically eliminated and the operation
of the proposed logic gate involves only in the metastable states of
the atom and hence is not affected by spontaneous emission.

This work was supported by NUS Research Grant No. R-144-000-071-305. X.L.F.
would also like to acknowledge the support of the National Natural Science
Foundation of China Grant No. 60578050.

\end{document}